\documentclass[aps,prd,twocolumn,nofootinbib,superscriptaddress]{revtex4}

\usepackage[utf8]{inputenc}	
\usepackage{amssymb}
\usepackage{amsmath}
\usepackage{graphicx}
\usepackage{subfigure}
\usepackage{tabularx}

\usepackage[dvips]{color} 
\begin{document}
	
	\title{QCD color superconductivity in compact stars: color-flavor locked quark star candidate for the gravitational-wave signal GW190814}
	\author{Zacharias Roupas}
	\email{Zacharias.Roupass@bue.edu.eg}
	\affiliation{Centre for Theoretical Physics, The British University in Egypt, Sherouk City 11837, Cairo, Egypt}
	
\author{Grigoris Panotopoulos}
\author{Il{\'i}dio Lopes}
\affiliation{Centro de Astrof{\'i}sica e Gravita{\c c}{\~a}o, Instituto Superior T{\'e}cnico-IST, Universidade de Lisboa-UL, Av. Rovisco Pais, 1049-001 Lisboa, Portugal.}

\begin{abstract}
	At sufficiently high densities and low temperatures matter is expected to behave as a degenerate Fermi gas of quarks forming { Cooper} pairs, namely a color superconductor, as was originally suggested by Alford, Rajagopal and Wilczek { {[}Nuclear Physics B 537, 443 (1999){]}}. The ground state is a superfluid, an electromagnetic insulator that breaks chiral symmetry, called the color-flavor locked phase. If such a phase occurs in the cores of compact stars, the maximum mass may exceed that of hadronic matter. The gravitational-wave signal GW190814 involves a compact object with mass $2.6{\rm M}_\odot$, within the so-called low mass gap. Since it is too heavy to be a neutron star and too light to be a black hole, its nature has not been identified with certainty yet. Here, we show not only that a color-flavor locked quark star with this mass is viable, but also we calculate the range of the model-parameters, namely the superconducting gap $\Delta$ and the bag constant $B$, that satisfies the strict LIGO constraints on the equation of state. We find that a color-flavor locked quark star with mass $2.6{\rm M}_\odot$ satisfies the observational constraints on the equation of state if $\Delta \geq 200{\rm MeV}$ and $B\geq 83{ \rm MeV}/{\rm fm^3}$ for a strange quark mass $m_s=95~{\rm MeV}/c^2$, and attains a radius $(12.7-13.6) {\rm km}$ and central density $(7.5-9.8) 10^{14}{\rm g}/{\rm cm}^3$.
\end{abstract}

\maketitle
	

\section{Introduction}

The LIGO/Virgo Collaboration announcement of the signal GW190814 \cite{2020ApJ...896L..44A} initiated a discussion on the nature of the merger's secondary component with mass $2.59^{+0.08}_{-0.09}{\rm M}_\odot$. There are theoretical and observational uncertainties regarding the maximum mass of neutron stars \cite{1996NuPhA.606..508M,PhysRevLett.32.324,Ozel_2012,Kiziltan_2013,Alsing_2018,PhysRevLett.121.161101,2019PhRvX...9a1001A,2020ApJ...896L..44A} and the lower mass of black holes \cite{Bailyn_1998,Ozel_2010,Farr_2011}, which do not allow concluding with certainty on the nature of this secondary GW190814 component.
The heaviest, observed, neutron star is $2.01\pm 0.04{\rm M}_\odot$ \cite{Antoniadis_2013} and PSR J0740+6620 may host a $2.14^{+0.10}_{-0.09} {\rm M}_\odot$ neutron star \cite{Cromartie_2019}, while stellar evolution seems to not allow stellar black holes to be formed with mass less than $5{\rm M}_\odot$ \cite{Bailyn_1998,Ozel_2010,Farr_2011}. Consequently, the existence and nature of compact objects in the mass range $\sim {[2.5,5]} {\rm M}_\odot$, the `low mass gap', are highly uncertain \cite{Bailyn_1998,Ozel_2010,Belczynski_2012}.

A primary candidate as the secondary GW190814 member is a stellar black hole \cite{2020ApJ...896L..44A,2020arXiv200703799F,2020arXiv200706057T,Sedrakian_2020}, but still multitude other proposals have been put forward. These include a
primordial black hole \cite{Vattis_2020,2020arXiv200700021L,2020arXiv200706481C,2020arXiv200703565J}, 
a heavy neutron star with stiff equation of state \cite{2020arXiv200616296T,2020arXiv200710999G,2020arXiv200910690D,2020arXiv200804491H,2020arXiv201002090B}, 
a fast pulsar \cite{2020arXiv201002090B,Zhang_2020,Most_2020,2020arXiv200705526T},
a compact star grown via accretion \cite{Safarzadeh_2020,2020ApJ...901L..34Y}, an anisotropic star \cite{2021Ap&SS.366....9R,Roupas_2020}, a spinning compact star with deconfinement phase transition \cite{2020arXiv200910731D}, a strange star \cite{2020arXiv201001509B}, a quark or hyperon star \cite{2020arXiv200708493D,2020arXiv200900942C,2020arXiv200907182Z}, a compact star with a population III star progenitor \cite{2020arXiv200713343K}, and modified gravity scenarios \cite{2020arXiv200810395N,2020arXiv200810884A}.

In the present work we propose to consider yet another possibility, namely the case of QCD color superconductivity. It has been shown by Alford, Rajagopal and Wilczek \cite{1999NuPhB.537..443A} (see also \cite{2001PhRvL..86.3492R,2008RvMP...80.1455A}) that at sufficiently high density, quarks of different color and flavour form { Cooper} pairs with the same Fermi momentum. This superfluid ground state was called a color-flavor locked (CFL) phase. { The quark matter in the CFL phase is electrically neutral and electrons cannot be present \cite{2001PhRvL..86.3492R}.} Later the CFL phase for up, down and strange quarks was argued to exist inside compact stars \cite{PhysRevD.102.043026,2004PhRvD..69e7505D,2009NuPhA.830..385A,2009MNRAS.400.1632K,2013PhRvD..88c4031W,2017PhRvC..95b5808F,2019EPJP..134..454L,2020IJMPD..2950044R}, called CFL quark stars. Studies of the structure of these objects have revealed that color superconductivity allows for large maximum masses \cite{2003A&A...403..173L,2004A&A...422L...1H,2017PhRvC..95b5808F}.
Using the constraints on the equation of state suggested by LIGO from the combined study of GW170817 and GW190814 \cite{2020ApJ...896L..44A} we will narrow down the CFL parameters, namely the superconducting gap $\Delta$ and the bag constant $B$, that allow for an equation of state of a $2.6{\rm M}_\odot$ CFL quark star that satisfies these constraints.

The plan of our work is the following. In the next section we briefly review the structure equations describing hydrostatic equilibrium of a CFL quark star interior solutions. In section \ref{sec:constraints} we calculate the region in the parameter space of a $2.6{\rm M}_\odot$ CFL quark stars for which the LIGO constraints are satisfied. Stability conditions are also discussed. Finally, in the last section we close with some concluding remarks.

\section{CFL quark stars}\label{sec:CFL_eos}

Lugones and Horvath \cite{2002PhRvD..66g4017L} have found that the CFL strange matter phase can be the true ground state of hadronic matter for a wide range of the parameters of the model (see also \cite{2017PhRvC..95b5808F}),  which are the QCD gap of { Cooper} pairs $\Delta$ and the bag energy density $B$ within an MIT bag model. They further derive a full equation of state and an analytic approximation, that is
\begin{align}
	P &= \frac{3\mu^4}{4\pi^2 \hbar^3 c^3} + \frac{9\alpha \mu^2}{2\pi^2 \hbar^3 c^3} - B,
	\\
	\rho &= \frac{9\mu^4}{4\pi^2 \hbar^3 c^5} + \frac{9\alpha \mu^2}{2\pi^2 \hbar^3 c^5} + \frac{B}{c^2},
\end{align}
where
\begin{equation}\label{eq:alpha}
	\alpha = -\frac{1}{6}m_s^2 c^4 + \frac{2}{3}\Delta^2
\end{equation}
and $m_s$ is the strange quark mass. From the above we get
\begin{align}
	\rho(P) &= 3\frac{P}{c^2} + 4\frac{B}{c^2} - 9\frac{\alpha (\mu (P))^2}{\pi^2 \hbar^3c^5} 
	\\
	(\mu(P))^2 &= -3\alpha+ \left( 9\alpha^2 + \frac{4}{3}\pi^2(P+B)\hbar^3c^3 \right)^{1/2}
\end{align}
that can be used to solve the Tolman-Oppenheimer-Volkoff (TOV) problem \cite{Tolman:1939jz,Oppenheimer:1939ne}
	\begin{align}
		\label{eq:TOV}
		\frac{dP}{dr} &= -\left(\rho(r) + \frac{P(r)}{c^2}\right)
		\frac{ \frac{G M(r)}{r^2} + 4\pi G \frac{P(r)}{c^2}r } 
		{ 1 - \frac{2G M(r)}{rc^2} },		\\
		\label{eq:massd}
		\frac{d M}{dr} &= 4\pi  \rho(r) r^2.
	\end{align}
We denote $M=M(r)$ the mass of a compact star included inside a radius $r$, and $P=P(r)$, $\rho=\rho(P(r))$ are the pressure and mass density, respectively. We shall denote $R$ the radius of the CFL-quark core and $M_{\rm QS} = M(r=R)$ the total mass. We integrate the TOV problem up to zero boundary pressure at the radius of the core and match with the exterior Schwarzschild metric \cite{Schwarzschild:1916uq}
\begin{equation}\label{exterior}
ds^2 = -f(r) dt^2 + f(r)^{-1} dr^2 + r^2 (d \theta^2 + sin^2 \theta d \phi^2)
\end{equation}
where the lapse function is given by
\begin{equation}\label{lapse}
f(r) = 1 - \frac{2 G M_{\rm QS}}{r c^2 },\; r \geq R .
\end{equation}
{ We fill further comment on the matching below, at the end of the section}.


\begin{table}[tb]
	\begin{center}
		\begin{tabular}{c c| c c}
			\multicolumn{2}{c|}{Models }
			&				
			\multicolumn{2}{c}{$M_{\rm QS}=2.6 {\rm M}_\odot$}
			\\
			\cline{1-4}
			& & & 
			\\ [-2.0ex]
			$\Delta {[{\rm MeV}]}$ &
			$B {[\frac{{\rm MeV}}{{\rm fm}^3}]}$ &
			$R {[{\rm km}]}$ &
			$\rho_0 {[10^{14}\frac{{\rm gr}}{{\rm cm}^3}]}$ 
			\\ [0.5ex]
			\cline{1-4}
			& & & 
			\\ [-2.0ex]
			$200$ & $83$ & $13.51$ & $8.97$ \\
			$210$ & $85$-$87$ & $13.62$-$13.41$ & $8.33$-$9.10$ \\
			$220$ & $89$-$91$ & $13.50$-$13.31$ & $8.49$-$9.24$ \\
			$230$ & $91$-$94$ & $13.56$-$13.30$ & $8.07$-$9.02$ \\
			$240$ & $95$-$98$ & $13.44$-$13.18$ & $8.29$-$9.22$ \\
			$250$ & $97$-$102$ & $13.47$-$13.07$ & $8.01$-$9.44$ \\
			$260$ & $99$-$105$ & $13.48$-$13.03$ & $7.81$-$9.36$ \\
			$270$ & $102$-$108$ & $13.41$-$12.98$ & $7.87$-$9.34$ \\
			$280$ & $104$-$112$ & $13.40$-$12.86$ & $7.76$-$9.65$ \\
			$290$ & $105$-$115$ & $13.45$-$12.80$ & $7.51$-$9.69$ \\
			$300$ & $108$-$118$ & $13.36$-$12.74$ & $7.65$-$9.78$ 
		\end{tabular}
		\caption{Bag constant, $B$, and semiconducting gap, $\Delta$, for $m_s=95{\rm MeV/c^2}$ that allow for a CFL quark star, as a GW190814-candidate, with $M_{\rm QS}=2.6{\rm M}_\odot$ which meets the LIGO constraints of Figure \ref{fig:EoS_match}. On the third and fourth columns are depicted the radius and central density values of the quark star for each model. These radius values correspond to compactness $0.28-0.30$.}
		\label{tab:models}
\end{center}\end{table}


The bag constant $B$ induces a mass density-scale, characteristic for the system, which in turn induces a radius-scale and a mass-scale
\begin{equation}
	\rho_\star \equiv \frac{B}{c^2},\quad 
	r_\star \equiv \left(\frac{4\pi G}{c^2}\rho_\star \right)^{-1/2},\quad 
	M_\star \equiv \frac{r_\star c^2}{G}.
\end{equation}
We further introduce the dimensionless variables
\begin{equation}
	x = \frac{r}{r_\star},\quad 
	\tilde{\rho} = \frac{\rho }{\rho_\star},\quad
	\tilde{P} = \frac{P}{\rho_\star c^2},\quad
	\tilde{M} = \frac{M}{M_\star}
\end{equation}
as well as the dimensionless quantities
\begin{equation}
	\lambda = \frac{9\alpha^2}{\pi^2 B \hbar^3 c^3},\quad
	\kappa (\tilde{P}) = \frac{\mu(\tilde{P})^2}{\alpha}.
\end{equation}
The problem is reformulated in the dimensionless format
	\begin{align}
	\label{eq:TOV_non_d}
	\frac{d\tilde{P}}{dx} &= -\left(\tilde{\rho}(x) + \tilde{P}(x)\right)
	\frac{ \frac{\tilde{M}(x)}{x^2} + \tilde{P}(x) x }{ 1 - \frac{2 \tilde{M}(x)}{x} },		\\
	\label{eq:massd_non_d}
	\frac{d \tilde{M}}{dx} &= \tilde{\rho}(x) x^2.
\end{align}
with
\begin{align}
	\tilde{\rho}(\tilde{P}) &= 3\tilde{P} + 4 - \lambda\cdot \kappa(\tilde{P})
	\\
	\kappa(\tilde{P}) &= -3 + 3\left( 1 + \frac{4}{3\lambda}(\tilde{P}+1) \right)^{1/2}
\end{align}
where
\begin{equation}\label{eq:lambda}
	\lambda = 0.118682\left(\frac{\alpha}{(100{\rm MeV})^2} \right)^2\left(\frac{B}{100{\rm MeV}/{\rm fm}^3} \right)^{-1}
\end{equation}
	and $\alpha$ is given in (\ref{eq:alpha}). We may also write
	\begin{align}
		\rho_\star &= 1.7827\cdot 10^{14} {\rm gr}/{\rm cm}^3
		\frac{B}{100{\rm MeV}/{\rm fm}^3}, \\
		r_\star &= 24.518{\rm km} \left(\frac{B}{100{\rm MeV}/{\rm fm}^3} \right)^{-1/2}, \\
		M_\star &= 16.5989{\rm M}_\odot \left(\frac{B}{100{\rm MeV}/{\rm fm}^3} \right)^{-1/2}.
	\end{align}
{ Due to the matching with the exterior Schwartzschild geometry the system (\ref{eq:TOV_non_d})-(\ref{eq:massd_non_d}) is subject to the constraint
\begin{equation}\label{eq:P_cond}
	\tilde{P}\left(\frac{R}{r_\star}\right) = 0,
\end{equation}
from which the radius of the CFL core may be calculated for any given central pressure. The mass is subject to the initial condition
\begin{equation}
	\tilde{M}(0) = 0,
\end{equation}
while following integration of (\ref{eq:TOV_non_d})-(\ref{eq:massd_non_d}) the total mass is given simply by the value $\tilde{M}(R/r_\star)$, where $R$ is such that (\ref{eq:P_cond}) is satisfied.
}

{ The matching with Schwartzschild geometry in the exterior would imply that the whole star is comprised of CFL quark matter, but this is not what we argue here to be the case. We consider this matching only as an approximation and to be precise the condition (\ref{eq:P_cond}) is realized as
	$
		P(R)/\rho_\star c^2 \ll 1 .
	$
	This approximation is sufficient for our purposes, that are to estimate the mass of the star for an inner core that satisfies the CFL quark matter equation of state and compare with the LIGO contraints in this high density region. The approximation is valid provided that the mass of the core dominates the mass of the star, as is commonly assumed (e.g. by Oppenheimer \& Volkoff \cite{Oppenheimer:1939ne}). 
	
	In a more realistic scenario, the hadronic matter would undergo a first-order phase transition designating QCD deconfinement close to the nuclear saturation density \cite{2020NatPh..16..907A}. In the case of CFL color superconductivity, as Sch\"afer \& Wilczek \cite{1999PhRvL..82.3956S} discussed, it is possible the symmetries of the CFL phase to be smoothly connected to those of a superfluid baryon phase in hadronic matter, without a sharp phase separation. In lack of sufficient observational evidence \cite{2020arXiv201008834D} and definite theoretical general consensus, we preferred not to match the CFL phase of the interior core with an exterior metric on the crust or outer core generated at some density by some hadronic equation of state, among a high multitude of possibilities. This would also delimit our main argument that focuses on the CFL equation of state in the core of the secondary GW190814 and not on the equation of state in the outer layers of the star. Instead, we preferred to apply the simple assumption that the quark core dominates the neutron star mass, which justifies a simple matching with Schwartzschild exterior geometry, following the tradition initiated by Oppenheimer \& Volkoff.}


\begin{figure}[tb]
	\begin{center}
		\subfigure[]{
			\label{fig:EoS_match}
			\includegraphics[scale = 0.5]{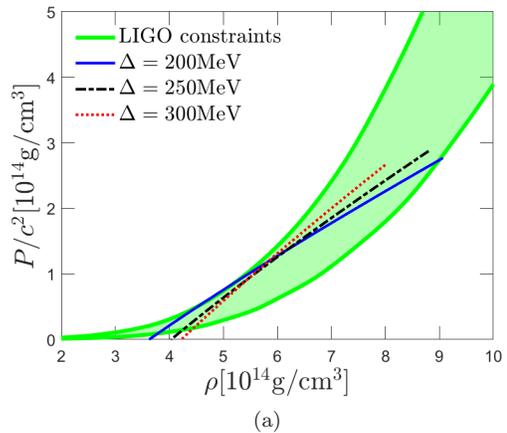}   }
		\subfigure[]{
			\label{fig:M-R_curves}
			\includegraphics[scale = 0.5]{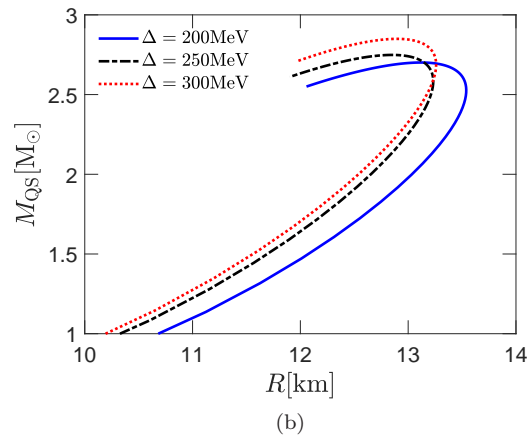}  }
		\caption{The solid blue, black dash-dotted, red dotted lines, correspond respectively to models  $\{\Delta = 200{\rm MeV},B=83{\rm MeV}/{\rm fm}^3\}$, $\{\Delta = 250{\rm MeV},B=100{\rm MeV}/{\rm fm}^3\}$, $\{\Delta = 300{\rm MeV},B=110{\rm MeV}/{\rm fm}^3\}$ with $m_s=95{\rm MeV/c^2}$. 
			(a) The equation of state for each solution $M_{\rm QS} = 2.6{\rm M}_\odot$ that satisfy LIGO constraints. 
			(b) The $M-R$ curves for the three models. The maximum masses and corresponding radii are for each case respectively $(2.70{\rm M}_\odot,13.03{\rm km})$, $(2.75{\rm M}_\odot,12.85{\rm km})$, $(2.85{\rm M}_\odot,12.90{\rm km})$.
		}
		\label{fig:M-R_curves+EoS_match}
	\end{center} 
\end{figure}


\begin{figure}[tb]
	\begin{center}
		\includegraphics[scale = 0.5]{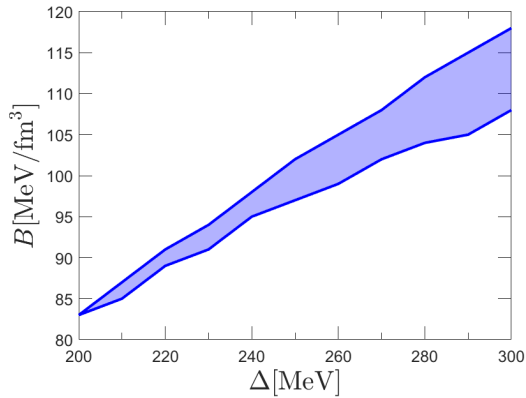} 
		\caption{ The range of values of the bag constant $B$ and semiconducting gap $\Delta$, assuming $m_s=95 ~{\rm MeV}/c^2$, for which a CFL quark star with mass $M_{\rm QS} = 2.6{\rm M}_\odot$ meets the LIGO constraints on the equation of state, depicted in Figure \ref{fig:EoS_match}.This is a visualization of Table \ref{tab:models}.
		}
		\label{fig:BvsDelta_match}
	\end{center} 
\end{figure}


\section{Analysis}\label{sec:constraints}

The green shaded region of Figure \ref{fig:EoS_match} designates the LIGO constraints (see Figure 8 of \cite{2020ApJ...896L..44A}) on the equation of state (EoS) imposed by the combined analysis of GW170817 and GW190814. Setting $m_s = 95 ~{\rm MeV}/c^2$, we calculate the range of parameter values for $\Delta,B$ for which these constraints are met, for a CFL quark star with $M_{\rm QS} = 2.6 ~{\rm M}_\odot$, as in Table \ref{tab:models}. { Our calculated values of $\Delta,B$ are the same at three digits accuracy for the whole range of strange mass $m_s = 93^{+11}_{-7} ~{\rm MeV}/c^2$ given in the latest review by the Particle Data Group \cite{Zyla:2020zbs}}.

 The radius of the corresponding CFL quark star is found to lie within the range $(12.7-13.6){\rm km}$, that corresponds to compactness values $C_{\rm QS} \equiv GM_{\rm QS}/Rc^2 = 0.28-0.30$. The CFL quark star is almost twice as compact than a { $1.4{\rm M}_\odot$} canonical neutron star satisfying the same LIGO constraints on the EoS, which attains compactness $C_{\rm NS} = 0.16$ { according to the LIGO/Virgo analysis} \cite{2020ApJ...896L..44A}.
 
	  Our values of $\Delta$, $B$ given in Table \ref{tab:models} are constrained with minimum values $\Delta \geq 200 {\rm MeV}$, $B \geq 83{\rm MeV}/{\rm fm}^3$. From the QCD side there have been made several estimates of $B$ depending for example on which energies to include in $B$ and which in quarks 
  	\cite{1974PhRvD..10.2599C,PhysRev.122.345,PhysRev.124.246,2005PhR...407..205B}. In particular, in the early MIT bag model it was estimated that $B\sim (60-80){\rm MeV}/{\rm fm}^3$ \cite{1974PhRvD..10.2599C}. On the other hand, in the Nambu-Jona-Lasinio (NJL) model \cite{PhysRev.122.345,PhysRev.124.246} it has been calculated $B \sim 296 {\rm MeV}/{\rm fm}^3$ \cite{2005PhR...407..205B}. 
 	Likewise, the value of quark pairing gap $\Delta $ depends on phenomenological assumptions. Within the context of the NJL model, for the densities of interest, it has been estimated that $\Delta \sim (100-200){\rm MeV}$
 	 \cite{2018RPPh...81e6902B} or higher, depending on assumptions regarding the, so called, vector repulsion parameter and the diquark pairing interaction parameter. Thus, our $\Delta$, $B$ values of Table \ref{tab:models} are consistent with the current phenomenology on the subject. 

 In Figure \ref{fig:EoS_match} we show for three concrete models that the LIGO constraints are satisfied for a CFL quark star with mass $M_{\rm QS} = 2.6{\rm M}_\odot$. The allowed region in the parameter space is then depicted graphically in Figure \ref{fig:BvsDelta_match}. Our results show that 
for a given energy gap $\Delta$, the bag constant $B$ takes values in a certain range, which becomes wider as $\Delta$ increases. { Note that the LIGO constraints naturally do not allow for exactly zero pressure at any density. That is why at very low pressures of Figure \ref{fig:EoS_match} the pressure of all models lie outside the green shaded region, since in our approximation of the star the exterior metric is matched with the Schwartzschild solution that requires zero pressure on the boundary. }

 In Figure \ref{fig:M-R_curves} we demonstrate for the same three models that the maximum mass exceeds $2.6 {\rm M}_{\odot}$, and therefore the CFL EoS considered here allows for a quark star with such high mass. { Note that the LIGO combined study of GW170817 and GW190814 gives for a canonical $1.4 {\rm M}_{\odot}$ neutron star a radius of $12.9^{+0.8}_{-0.7} {\rm km}$ \cite{2020ApJ...896L..44A}. In our models, we get the larger radius, $R = 11.8 {\rm km}$, of an $1.4 {\rm M}_{\odot}$ CFL quark core, for the lower $\Delta$, $B$ values, namely $\Delta = 200{\rm MeV}$, $B=83{\rm MeV}/{\rm fm}^3$. However, it was shown by Annala et al. \cite{2020NatPh..16..907A} that neutron stars of different masses have cores with strikingly different properties and in particular that an $1.4M_{\odot}$ neutron star cannot contain a quark core.}
 
 { Regarding tidal deformability, we remark that there was no evidence of measurable tidal effects in the signal GW190814 as reported by the LIGO/Virgo collaboration \cite{2020ApJ...896L..44A} and therefore the tidal deformability of the secondary GW190814 component was not estimated by LIGO/Virgo. For the three indicative models of Figure \ref{fig:M-R_curves+EoS_match}, namely $\{\Delta, B\} = \{200{\rm MeV}$, $83{\rm MeV}/{\rm fm}^3\}$, $\{250{\rm MeV},100{\rm MeV}/{\rm fm}^3\}$, $\{300{\rm MeV},110{\rm MeV}/{\rm fm}^3\}$, the tidal deformability $\Lambda$ (for a definition see for instance \cite{Hinderer,Damour,Lattimer}) of a $2.6{\rm M}_\odot$ CFL quark core is respectively $\Lambda =  5.22$, $4.51$, $4.56$. }

\begin{figure*}[tb]
	\begin{center}
		\subfigure[]{
			\label{fig:S_condition}
			\includegraphics[scale=0.38]{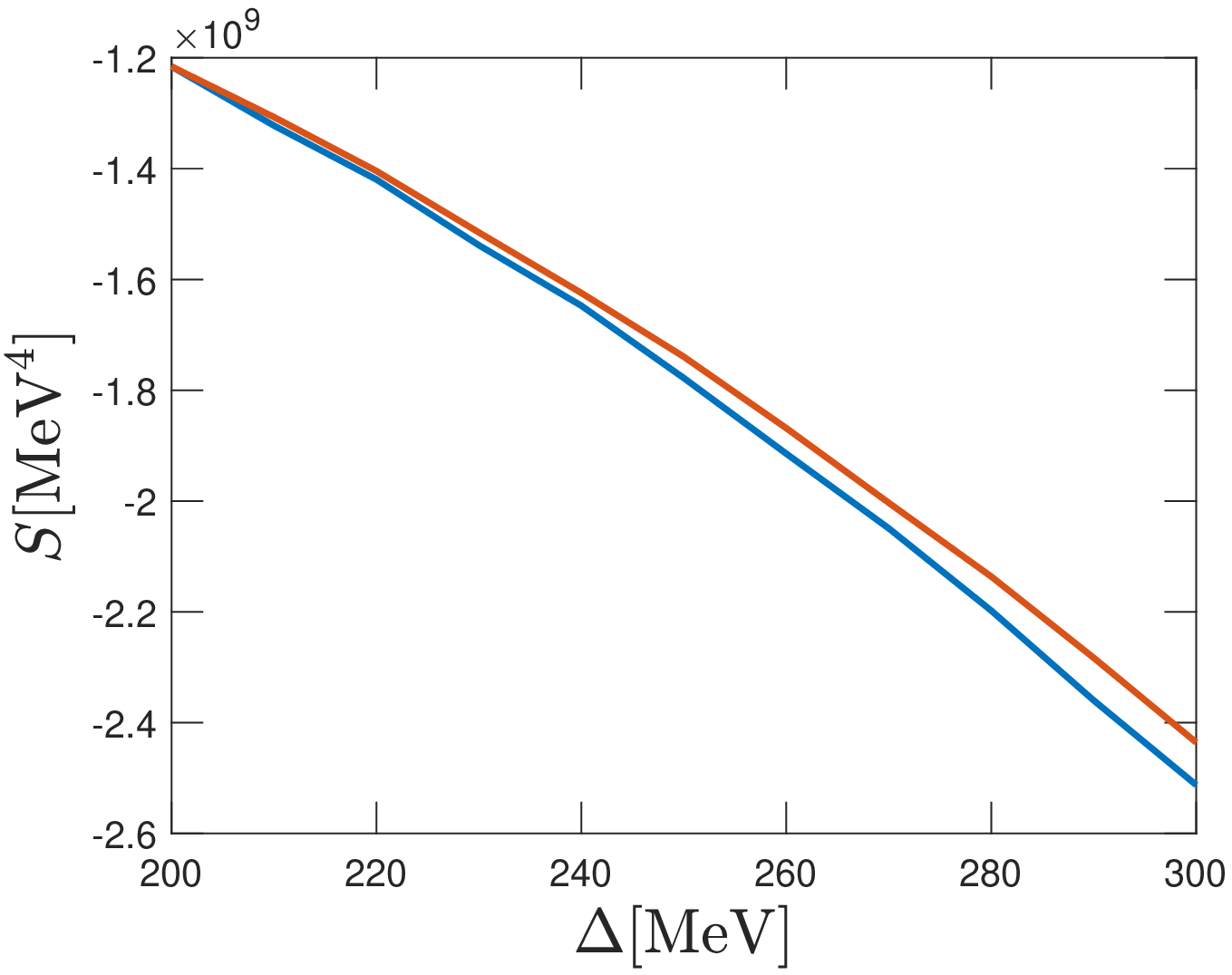} }
		\subfigure[]{
			\label{fig:adiabatic_index}
			\includegraphics[scale = 0.38]{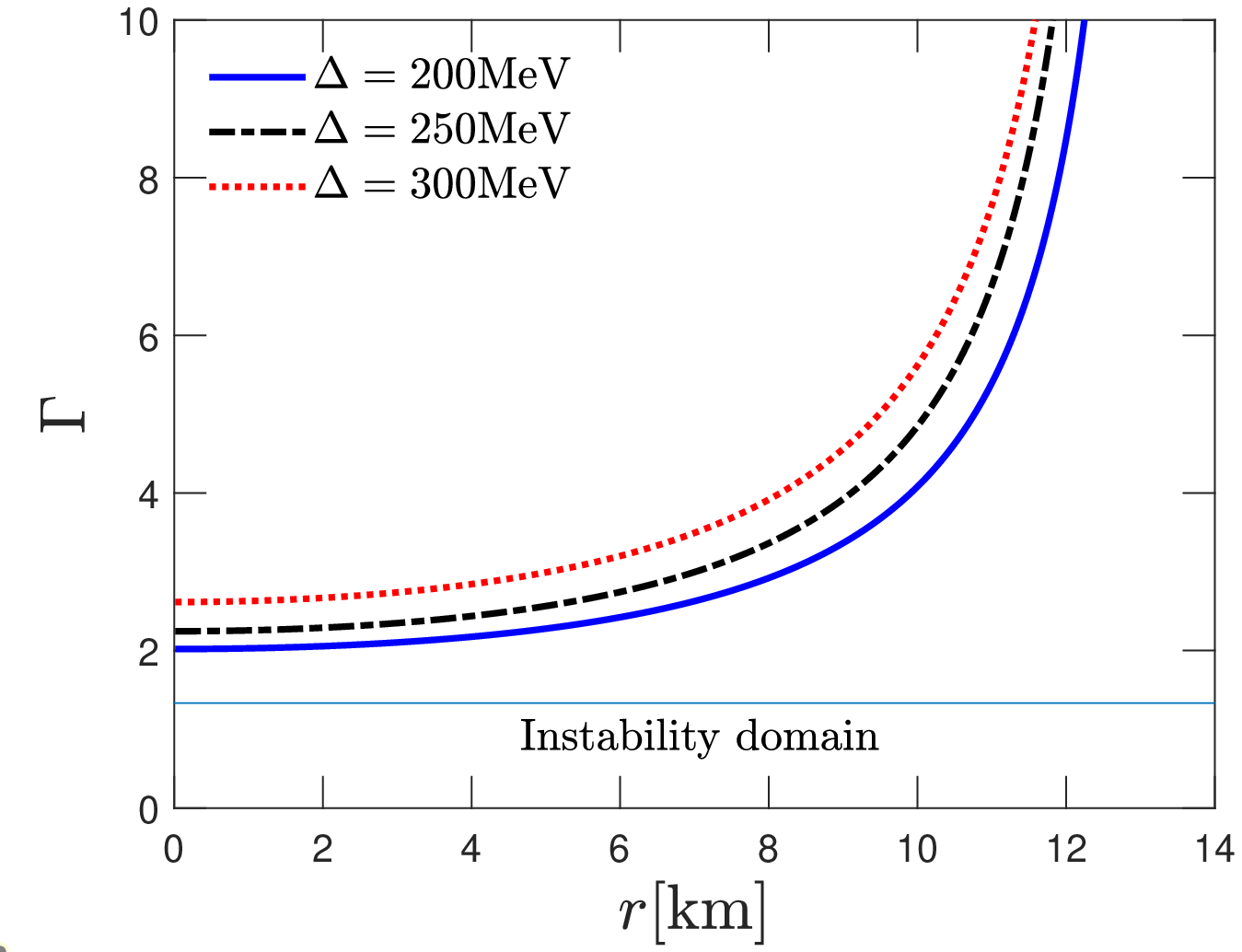}   }
		\subfigure[]{
			\label{fig:v_s_square}
			\includegraphics[scale = 0.38]{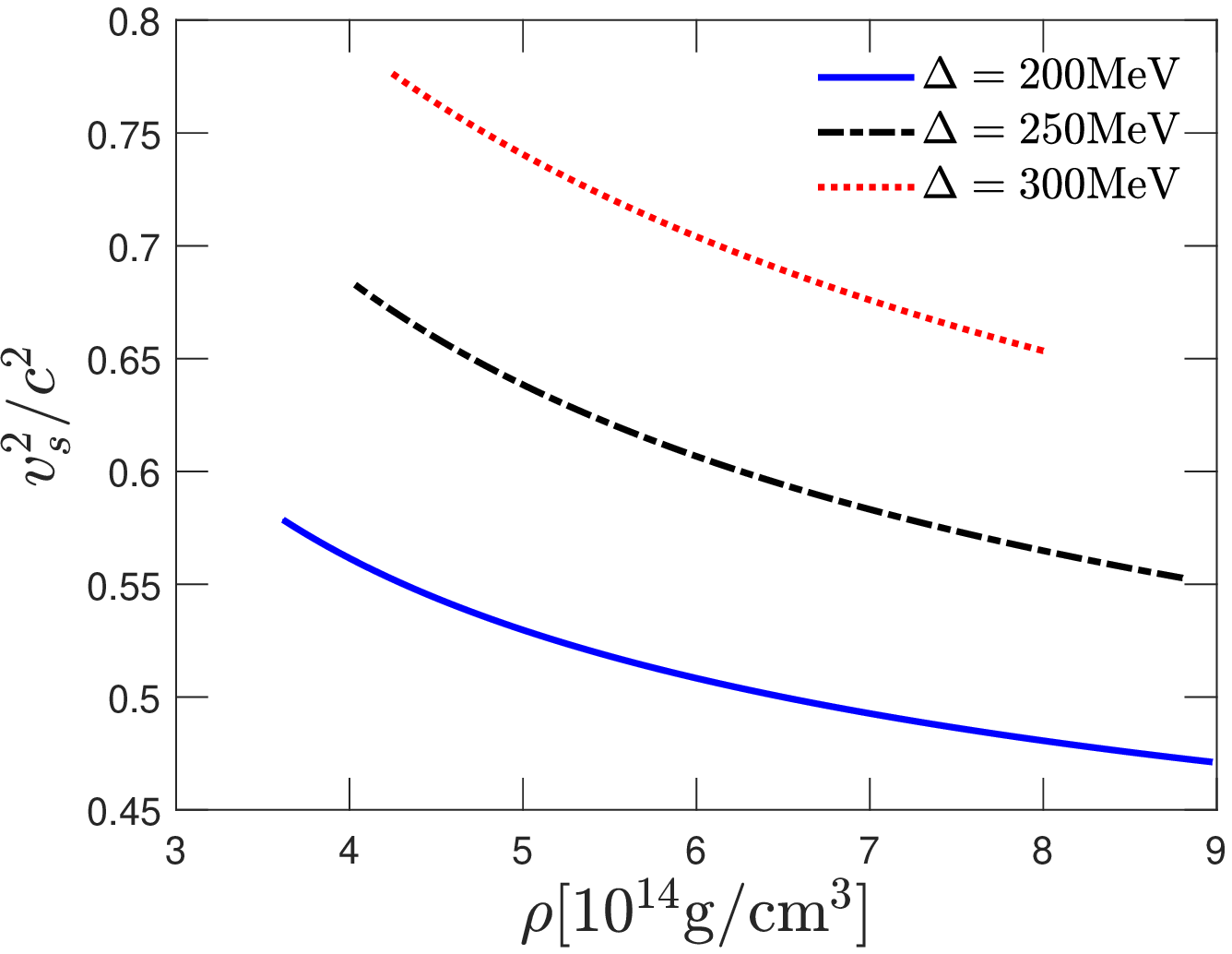}  }
		\caption{ Stability and physical requirements. (a) The quantity $S$ defined in (\ref{eq:S}) with respect to the superconducting gap $\Delta$ for the marginal values of  $B$ for all models of Table \ref{tab:models}. Stability of the CFL phase requires $S<0$.  
		(b) The adiabatic index $\Gamma$ with respect to the radius in the interior of a $2.6{\rm M}_\odot$ CFL quark star for the three models of Figure \ref{fig:M-R_curves+EoS_match}. Stability requires $\Gamma > 4/3$.
		(c) The velocity of sound  squared $v_s^2$ with respect to the density in the interior of a $2.6{\rm M}_\odot$ CFL quark star for the three models of Figure \ref{fig:M-R_curves+EoS_match}. Causality requires $v_s < c$.
		}
		\label{fig:stability_conditions}
	\end{center} 
\end{figure*}

We further study stability and physical requirements for the solutions of Table \ref{tab:models}. Firstly, we stress that CFL quark matter is absolutely stable if the energy per baryon is smaller than the neutron mass, $m_n = 939~{\rm MeV/c^2}$. This condition is satisfied if \cite{2017PhRvC..95b5808F}
\begin{equation}\label{eq:S}
S \equiv B\hbar^3c^3 + \frac{m_s^2m_n^2 c^8}{12\pi^2} - \frac{\Delta^2 m_n^2 c^4}{3\pi^2}  - \frac{m_n^4 c^8}{108\pi^2} < 0.   
\end{equation}
In Figure \ref{fig:S_condition} we demonstrate that this stability condition is satisfied for all parameter values of Table \ref{tab:models}.


In addition, the $2.6{\rm M}_\odot$ solution satisfies the stability condition of the adiabatic index \cite{Moustakidis:2016ndw} 
\begin{equation}
	\Gamma \equiv \frac{1}{c^2} \frac{d P}{d \rho} \left[ 1 + \frac{\rho c^2}{P} \right] > \frac{4}{3},
\end{equation}
for all models of Table \ref{tab:models}. We demonstrate this in Figure \ref{fig:adiabatic_index} for the three models of Figure \ref{fig:M-R_curves+EoS_match}.
For the same three models we demonstrate in Figure \ref{fig:v_s_square} that the causality condition
\begin{equation}
0 < v_s^2 \equiv \frac{d P}{d \rho} < c^2
\end{equation}
is also satisfied, where $v_s$ denotes the speed of sound.

{ It is evident that the speed of sound violates the conformal limit, which would require $v_s^2 \leq c^2/3$, in accordance with earlier calculations of the speed of sound in deconfined quark matter \cite{2015PhRvL.114c1103B, 2018ApJ...860..149T}. In particular, at the phase transition point to quark deconfinement the speed of sound presents a pick way above the conformal limit and then it decreases with increasing density, in accordance with Figure \ref{fig:v_s_square}. The phase transition is predicted \cite{2020arXiv200616296T} to occur at about $(1-3)\rho_{\rm sat}$, where $\rho_{\rm sat} = 2.7{\rm g}/{\rm cm}^3$ is the nuclear saturation density in direct agreement with our Figure \ref{fig:v_s_square}, where the phase transition point corresponds to the lower density of each curve.}
 
{ Note that the value of $v_s^2$ depends on $\alpha$ defined in (\ref{eq:alpha}). In the conformal limit, i.e. when all mass scales are set to zero, $\alpha = m_s = \Delta = 0$, it is $v_s^2 = c^2/3$. For $\alpha < 0$, it is $v_s^2 < c^2/3$, while for $\alpha > 0$ it is $v_s^2 > c^2/3$. It is a non-vanishing and sufficiently large $\Delta$ that forces $v_s$ to violate the conformal limit. In all our models $\alpha$ is positive, however there are viable CFL models for which $\alpha$ is negative, see e.g \cite{2017PhRvC..95b5808F}.} 
 
Finally, we require \cite{Deb:2015vda,Deb:2016lvi,Panotopoulos:2020zqa,Panotopoulos:2020uvq} that the strong energy condition
\begin{equation}
	\rho c^2 + P  \geq  0\,, \quad \rho c^2 + 3P \geq 0\,
\end{equation} 
is satisfied. This is true for all models of Table \ref{tab:models}, as it is evident in Figure \ref{fig:EoS_match} for our three indicative models. Note that since the strong energy condition is satisfied it follows that all other energy conditions are satisfied, namely 
the weak energy condition $\{\rho \geq 0, \; \rho c^2+ P \geq 0\}$, the null energy condition $\{\rho c^2 + P  \geq  0\}$ and the dominant energy condition $\{ \rho c^2 \geq \lvert P \rvert\}$.

\section{Conclusions}\label{sec:conclusions}

In the present work we have investigated the possibility that strange quark stars in the CFL phase comprise the secondary GW190814 component, with an observed mass at $M_{\rm QS}=2.6 ~{\rm M}_\odot$. QCD superconductivity effects lead to a non-linear, but still analytical, EoS characterized by three parameters, namely the bag constant $B$, the superconducting energy gap $\Delta$ as well as the mass of the strange quark $m_s$. Assuming for the latter a numerical value compatible with the Particle Data Group review at $m_s = 95 ~{\rm MeV}/c^2$, we obtain the region in the $(B-\Delta)$ plane for which the stringent LIGO constraints on the EoS are met. In particular, we have obtained the $M-R$ relationships, and our main numerical results show that the EoS adopted here can support a CFL quark star with a mass $M_{\rm QS}=2.6 ~{\rm M}_\odot$ provided that the semiconducting gap $\Delta \geq 200{\rm MeV}$, and that the bag constant $B\geq 83{\rm MeV}/{\rm fm^3}$ with a range of values depicted in Figure \ref{fig:BvsDelta_match}. For these values, we further verify that stability, causality and energy conditions are met, suggesting that our solutions are physical within the context of General Relativity. The radius of the corresponding CFL quark star is found to lie within the range $(12.7-13.6){\rm km}$, while the central density $\rho_0$ is found to take values in the range $(7.5-9.8)10^{14}{\rm g}/{\rm cm^3}$.


\section*{Acknowlegements}

The authors G.~P. and I.~L. thank the Fun\-da\c c\~ao para a Ci\^encia e Tecnologia (FCT), 
Portugal, for the financial support to the Center for Astrophysics and Gravitation-CENTRA, 
Instituto Superior T\'ecnico, Universidade de Lisboa, through the Project No.~UIDB/00099/2020 
and No.~PTDC/FIS-AST/28920/2017.


\bibliography{CFL_quark_star_GW190814}

\end{document}